\documentclass[preprint,3p,twocolumn,12pt]{elsarticle}
\usepackage{amssymb}
\usepackage{amsmath}
\usepackage{slashbox}

\journal{Astroparticle Physics}

\begin{document}

\begin{frontmatter}



\title{Particle Spectra from Acceleration at Forward and Reverse Shocks of Young Type Ia Supernova
Remnants}


\author[lab1]{I.Telezhinsky}
\author[lab3]{V.Dwarkadas}
\author[lab1,lab2]{M.Pohl}

\address[lab1]{DESY, Platanenallee 6, 15738 Zeuthen, Germany}
\address[lab3]{University of Chicago, Department of Astronomy \& Astrophysics,
5640 S Ellis Ave, AAC 010c, Chicago, IL 60637, U.S.A.}
\address[lab2]{Universit\"{a}t Potsdam, Institut f\"{u}r Physik \& Astronomie,
Karl-Liebknecht-Strasse 24/25, 14476 Potsdam, Germany}

\begin{abstract}
We study cosmic-ray acceleration in young Type Ia Supernova Remnants
(SNRs) by means of test-particle diffusive shock acceleration theory
and 1-D hydrodynamical simulations of their evolution. In addition to
acceleration at the forward shock, we explore the particle 
acceleration at the reverse shock in the presence of a possible substantial magnetic field, and consequently the impact of this acceleration on the particle spectra in the remnant. We investigate
the time evolution of the spectra for various time-dependent profiles
of the magnetic field in the shocked region of the remnant. We test a
possible influence on particle spectra of the Alfv\'enic drift of
scattering centers in the precursor regions of the shocks. In
addition, we study the radiation spectra and morphology in a broad
band from radio to gamma-rays. It is demonstrated that the reverse
shock contribution to the cosmic-ray particle
population of young Type Ia SNRs may be significant, modifying the spatial
distribution of particles and noticeably affecting the
volume-integrated particle spectra in young SNRs. In particular spectral structures
may arise in test-particle calculations that are often discussed as
signatures of non-linear cosmic-ray modification of shocks. Therefore,
the spectrum and morphology of emission, and their time evolution, 
differ from pure forward-shock solutions.

\end{abstract}


\end{frontmatter}

% \linenumbers

\section{Introduction}
\label{intro}

The energy density in local cosmic rays (CRs), when extrapolated to
the whole Galaxy, implies the existence of powerful accelerators
inside the Galaxy.  Supernova Remnants (SNRs) have long been thought
to be the main candidates on account of energetics
\cite{GinSyr61}. Later, diffusive shock acceleration (DSA) of charged
particles at the SNR shocks was considered \cite{Axfetal77, Kry77,
  Bel78, BlaOst78} and is now thought to be the most likely mechanism
that can accelerate CRs up to $10^{15}$~eV, where the cosmic-ray
spectrum shows a break known as the 'knee' \cite{LagCes83}. If they
are very efficiently accelerated, CRs can produce a feedback on the
plasma flow in SNRs through their pressure, and can modify the shock
structure by decelerating the incoming plasma flow when streaming
through it in the precursor region (see \cite{MalDru01} for a review
and references therein).  Observationally, it is unclear whether or
not the CR-acceleration efficiency is sufficiently high for a
significant feedback \cite{2005ApJ...634..376W,2011arXiv1103.3211R},
but if so, SNR shocks may become modified, and the particle spectra
may deviate from the classical power-law slope of $s=2$ found in
test-particle calculations.  The spectra produced by modified shocks
show concave structure with softer ($s>2$) low-energy and harder
($s<2$) high-energy parts.

The CR spectrum observed at Earth suggests that sources produce softer
spectra than predicted by non-linear DSA (NDSA) theory
\cite{Aveetal09}. If the CR sources produce very hard spectra, the
energy dependence of the diffusion coefficient in the Galaxy needs to
be much stronger than is allowed by the observed CR anisotropy
\cite{Ptuetal06}. It should be noted, though, that the CR source
spectrum is that of the CR escaping from their sources, the time
integral of which could be considerably softer than the spectrum of
particles contained in the sources at a given time
\cite{2005A&A...429..755P}.  On the other hand, recent $\gamma$-ray
observations of some shell-type SNRs tend to show soft spectra
($\Gamma>2$): RX~J1713.7-3946 \cite{Ahaetal04, Ahaetal07a},
RX~J0852.0-4622 \cite{Ahaetal07b}, RCW~86 \cite{Ahaetal09}, SN~1006
\cite{Aceetal2010_1006}, Cas~A \cite{2010ApJ...714..163A,
  Abdetal10_CasA}, IC~443 \cite{2009ApJ...698L.133A, Abdetal10_ic443}
and other \cite{CasSla10, Abdetal10_w44, Abdetal10_w28}. This means
that the particle spectra, assuming the radiation was produced by
hadronic scenario, is also softer than NDSA predictions.

It is known that CR particles are scattered by magnetohydrodynamic
(MHD) waves.  These waves, alongside amplified magnetic fields
  (MFs), may be generated by instabilities invoked by CR streaming,
  which increases MHD turbulence \cite{Bel78, Bel04, ZirPtu08a,
    AmaBla09}, or directly by pure MHD turbulence induced by
  propagation of collisionless shock in media \cite{Che77, GiaJok07,
    Beretal09}.

Gyro-resonant instabilities favor the generation of MHD waves that
propagate with Alfv\'en velocity \cite{Ski75}, whereas non-resonant
instabilities can initially favor non-oscillating and non-propagating
modes. If the former dominate, the scattering centers move relative to
the plasma, and the inclusion of this so-called Alfv\'enic drift may
soften the spectra so that they could match experimental data
\cite{ZirPtu08b, Ptuetal10, KanRyu10, Capetal09a,
  Capetal11}. Alternatively, the bulk of Galactic CRs may be produced
in a low-efficiency or test-particle mode transferring only $\lesssim
10\%$ of the SNR energy to CRs \cite{KanRyu10}. This number is roughly
that needed to explain the origin of CRs by acceleration in Supernova
Remnants \cite{Rey08}.

Given the complexities faced by NDSA theory, in this paper we
  propose to investigate the diversity of particle spectra that can
be produced by young SNR shocks in a test-particle mode, concentrating
for simplicity on Type Ia SNRs in this paper. We consider acceleration
by both forward and reverse shocks in evolving
  spherically-symmetric systems. Many pieces of observational
    evidence suggest that the reverse shock (RS) may also accelerate
    particles to high energies. This includes the detection of
  non-thermal X-rays from the RS of Cas~A \cite{Gotetal01, HelVin08},
  1E~0102.2-7219 \cite{Sasetal06}, and RCW~86 \cite{Rhoetal02}, as
  well as radio emission from the RS of Kepler's SNR \cite{Deletal02}.
  CR acceleration at the RS is also supported by theory \cite{Elletal05, ZirAha10}. In any case, high-energy particles produced by the FS may be re-accelerated at the RS.
  \cite{Schetal10}. The preponderance of observational and
    experimental evidence, combined with earlier theoretical
    implications, suggests that it is essential to at least consider
    the effects of both shocks in accelerating particles. The total
  volume-integrated spectra in such a case will deviate from
expectations based on plane-parallel shocks and steady-state systems
employing the FS only.

The MF in the shocked interaction region is essential to
  determining the particle spectra, but is extremely hard to deduce
  observationally. This has led to the parameterization of the field
  in different models, which generally involves tying the field to
  dynamically or kinematically determinable parameters. Our intent is
  not to study the amplification of the field itself, which is beyond
  the scope of this paper, but to investigate the particle spectra
  that are produced under widely used assumptions of the field in the
  literature. We also explore how the inclusion of Alfv\'enic drift
affects the total particle spectra in SNR. Given the observational
data on MF amplification \cite{2011arXiv1102.3871A, Voeetal05} as well
as theoretical \cite{Bel78, Bel04} and numerical \cite{BelLuc01,
  ZirPtu08a} results, we consider two cases: a moderate (75$\mu$G) and
a strongly amplified (300$\mu$G) MF. Besides calculating the total
particle spectra, we produce wide-band spectral-energy distributions
(SEDs), and provide surface brightness profiles in the radio, X-ray
and $\gamma$-ray bands.

This work intends to improve upon the present state of the field in many ways (1) We take into account the free expansion phase of the expanding SNR, when the shock velocities are the highest, whereas most (though certainly not all) analyses have totally neglected the initial phase, and started directly with the Sedov phase. (2)  Our technique to solve the transport equation, and the corresponding high resolution spherically-symmetric simulations used, guarantee a very high level of accuracy and therefore an accurate rendering of the particle spectra for the chosen MF and diffusion coefficient. (3) We have taken into account the acceleration of CRs at the reverse shock, as suggested by recent observations. We have shown what observable features in SNRs we may expect if a non-negligible MF permits CR acceleration at the RS to high energies. This has not been explored in such detail before, including spherical symmetry and fully time-dependent cosmic-ray transport. Earlier studies \cite{Elletal05} investigated CR acceleration at the RS using an approximate, analytical model \cite{BerEll99} and assumed there is no MF amplification at the RS. In particular, they focused on non-linear effects arising from a high injection efficiency in low MF. Other authors \cite{ZirAha10} concentrated on the study of a specific object, which most probably is a core-collapse SNR, and have not made any general conclusions regarding type-Ia SNRs. (4) Our results therefore present a more complete study of the particle spectra and high-energy emission that is to be expected from type-Ia SNe with moderate acceleration efficiency.

\section{Cosmic-Ray Acceleration}
\label{modcr}

We consider a test-particle approach to diffusive shock acceleration
of CRs in SNRs \cite{Axfetal77, Kry77, Bel78, BlaOst78}. It has been
recently \cite{KanRyu10} shown that a test-particle description is
applicable if the CR pressure at the shock is less than 10\% of the
shock ram pressure.  Therefore, if one limits the amount of energy
contained in CRs and the CR pressure at the shock, the acceleration
can be treated by independently solving the cosmic-ray transport
equation and the hydrodynamic equations of SNR evolution. The
cosmic-ray transport equation is a diffusion-advection equation in
both space and momentum \cite{Ski75}:
\begin{align}
\frac{\partial N}{\partial t} = \nabla (D_r\, \nabla N -\vec v\, N) - \nonumber
\\ 
\frac{\partial}{\partial p} \left((N\,\dot{p})- 
\frac{\nabla \vec v}{3} \,N\,p  \right) + Q %\nonumber\\ 
%\frac{\partial}{\partial p} \left(p^2\, D_p
%\frac{\partial}{\partial p} \frac{N}{p^2} \right) + Q %\nonumber
\label{traneq}
\end{align}
where $N$ is the differential number density of cosmic rays, $D_r$ is
the spatial diffusion coefficient, $\vec v$ is the advective velocity
given by a 1-D hydrodynamical simulation, $\dot{p}$ are the energy
losses, 
%$D_p$ is the momentum-space diffusion coefficient, 
and $Q$ is
the source term representing the injection of the thermal particles
into the acceleration process given as
\begin{equation}
Q = \eta_{i} n_{u} |V_{sh} - v_{u}| \delta(r - R_{sh}) \delta(p - p_{inj}), 
\label{src}
\end{equation}
where $\eta_{i}$ is the injection efficiency parameter, $n_{u}$ is the number
density of plasma in the shock upstream region, $V_{sh}$ is the shock speed,
$v_{u}$ is the plasma velocity in the shock upstream region, $r$ is the distance
from the SNR center, $R_{sh}$ is the radius of the shock, $p$ is the particle
momentum, and $p_{inj}$ is the momentum of the injected particles. 

We assume the thermal leakage injection model
  \cite{Blaetal05}, where only the particles with momentum $p_{inj} >
  \psi p_{th}$ can be accelerated, with $\psi$ being a multiple of the
  particle thermal momentum, $p_{th}$. The efficiency of injection is
  determined as
\begin{equation}
\eta_i = \frac{4}{3 \sqrt{\pi}} (\sigma -1) \psi^3 e^{-\psi^2}
\end{equation}
where $\sigma$ is the shock compression ratio. Although $\sigma$ and
$p_{th}$ are obtained from hydro simulations, $\psi$ remains a free
parameter in the model. In our calculations we assume $\psi \simeq
4.45$, keeping $\eta_i \simeq 5 \times 10^{-7}$ sufficiently low
  to stay within the framework of the test-particle approximation. We
  note that since we are not interested here in the absolute ratio of
  electron to proton spectra, and lacking secure knowledge of the
  details of electron injection, as well as the time-scale of
thermalization between electrons and protons, we assume that electrons
and protons are injected equally and are at the same temperature,
i.e. the injected electron to proton ratio, $K_{e/p,i} = 1$. Thus,
$K_{e/p}$ obtained in our simulations is purely a result of the electron-to-proton mass ratio and corrected for standard ISM abundances.

We impose a free-escape spatial boundary upstream of the FS that
should account for escape of the highest-energy particles from the
system. We assume that all particles crossing the boundary leave the
system, therefore we set the CR number density, $N$, to zero at the
free-escape boundary, at $R_{\rm esc}=2R_{FS}$, where $R_{FS}$ is the
FS radius.

We also assume that the highest energy particles generate MHD waves in
the shock upstream region, and thus may amplify the MF \cite{Bel04,
  ZirPtu08a, AmaBla09, Capetal09a}.  Being scattered by MHD waves, the
CR particles undergo Bohm diffusion everywhere inside the escape
boundary \cite{Revetal09}. Also, the advective velocity in
Eq.~\ref{traneq}, given the presence of amplified MF, may include the
Alfv\'enic drift of scattering centers and thus may not coincide with
the plasma flow velocity \cite{ZirPtu08b, Capetal09a, KanRyu10,
  Ptuetal10}.

In the current paper we assume spherical symmetry and do not consider
the momentum-diffusion term representing stochastic (second order
Fermi) acceleration.

The main difficulty for the numerical solution of Eq.~\ref{traneq} is
that the diffusion coefficient $D_r$ is strongly dependent on the
particle momentum ($D_r = D_r(p) \sim p$, in case of Bohm diffusion)
and therefore spans a wide range of magnitudes, $\ln(p_{\rm
  max}/p_{inj})\approx 25$.  Formally, the spatial grid must resolve
the smallest diffusion length defined as $D_r(p_{inj})/V_{sh}$,
otherwise the modeled acceleration is artificially slow.  The
lowest-energy particles have a very small diffusion length, and thus a
uniform grid must contain millions (or even billions) of cells which
is computationally impossible.

Several approaches are known to date to overcome this numerical
difficulty. The so-called normalized grid is used in \cite{Beretal94},
where a substitution of variables is performed and the spatial
coordinates are normalized to the diffusion scale. Another approach
presented by \cite{KanJon06} consists of introducing a co-moving
frame, in which the shock is stationary, coupled with a mesh
refinement - a fine discretization is used only near the shock region,
where the low-energy particles are injected.

Here we use the second approach. We normalize the spatial coordinate
in Eq.~\ref{traneq} to the shock radius, introducing coordinate
$x=r/R_{sh}$, and then substitute the coordinate $x$ with a new
coordinate, $x_{\ast}$, for which we use a uniform grid when solving
Eq.~\ref{traneq}:
\begin{equation}
(x - 1)=(x_{\ast} - 1)^3
\label{coord}
\end{equation}
with Jacobian
\begin{equation}
\frac{dx}{dx_{\ast}} = 3(x_{\ast} - 1)^2
\end{equation}

These simple transformations allow us i) to place the shock at the
center of a known cell at any given time and ii) to resolve the shock
vicinity for low-energy particles with only a few hundred bins in the
spatial coordinate, $x_{\ast}$.  We use parametrizations for the
radial dependence of the MF as described in Section~\ref{modmf}, and
we obtain the hydrodynamical parameters (the plasma flow velocity,
density and the shock speed) needed to solve Eq.~\ref{traneq} from 1-D
hydrodynamic simulations of the SNR evolution described in
Section~\ref{modhd}.

Then, Eq.~\ref{traneq} is solved in spherically-symmetric geometry
using implicit finite-difference methods implemented in the
\textit{FiPy} \citep{fipy} library modules. We make separate runs for
the FS and the RS, using the appropriate coordinate transformations to
obtain the highest resolution where injection occurs. Then we sum the
results because they are linear and independent. Some of the
highest-energy particles may reach the other shock and be
re-accelerated there. However, the energy of these particles is
already so high, and their number so low, that the poor resolution at
the other shock does not visibly affect the results. Finally,
  we check the CR-to-ram pressure ratio as well as energy containment
  in CRs \textit{a posteriori}. In our calculations these values are
  never allowed to exceed 10\%.

\section{Hydrodynamics}
\label{modhd}

To study the evolution of the SNR shock waves over time, we
have performed spherically-symmetric hydrodynamic simulations
using the VH-1 code \cite{BloLun93}, a
3-dimensional hydrodynamic code based on the Piecewise Parabolic
Method \cite{colwoo84}. The simulations are based on an expanding grid
\cite{Dwa05, Dwa07}, which tracks the motion of the outer shock and
expands along with it. Thus, a high resolution is maintained right from
the start of the simulations, which is essential considering that the
remnant size increases by about 6 orders of magnitude during the
simulation. The expanding grid also has another advantage - the
position of the forward shock is almost stationary on the grid over
most of the evolution, which is useful in computing the particle
spectrum. Furthermore, the grid expansion is adjusted such that, as
far as possible, the shocked interaction region between the inner and
outer shocks occupies most of the grid. Although cooling of the
material is incorporated into the simulations via a cooling function,
the velocities and densities are such that cooling is not effective
and does not play a role.

In order to study the evolution of young SNe in the ambient medium, we
need at minimum two parameters - a description of the density profile
of the material ejected in the explosion, and that of the medium into
which the SN is expanding. In this first paper, we consider the
evolution of a Type Ia SN, which are thought to arise from the
thermonuclear deflagration of a white dwarf. Since the progenitor star
is a low-mass star and (presumably) does not undergo significant
mass-loss, the medium surrounding it is assumed to be a constant-density
interstellar medium.

The ejecta density structure is more uncertain. However, by comparing
spherically-symmetric models of Type Ia SN explosions, it was
found \cite{DwaChe98} that the ejecta structure of a Type Ia SN after explosion can
best be represented by an exponential ejecta density profile. This is
the profile that we have used in our simulations. Since an exponential
introduces an additional dimensional parameter to satisfy a
non-dimensional exponent, the resulting SN evolution is no longer
self-similar. The SN is homologously expanding, therefore the ejecta
velocity simply increases linearly with radius.

\begin{figure}[!t]
\begin{center}
\includegraphics[width=0.49\textwidth]{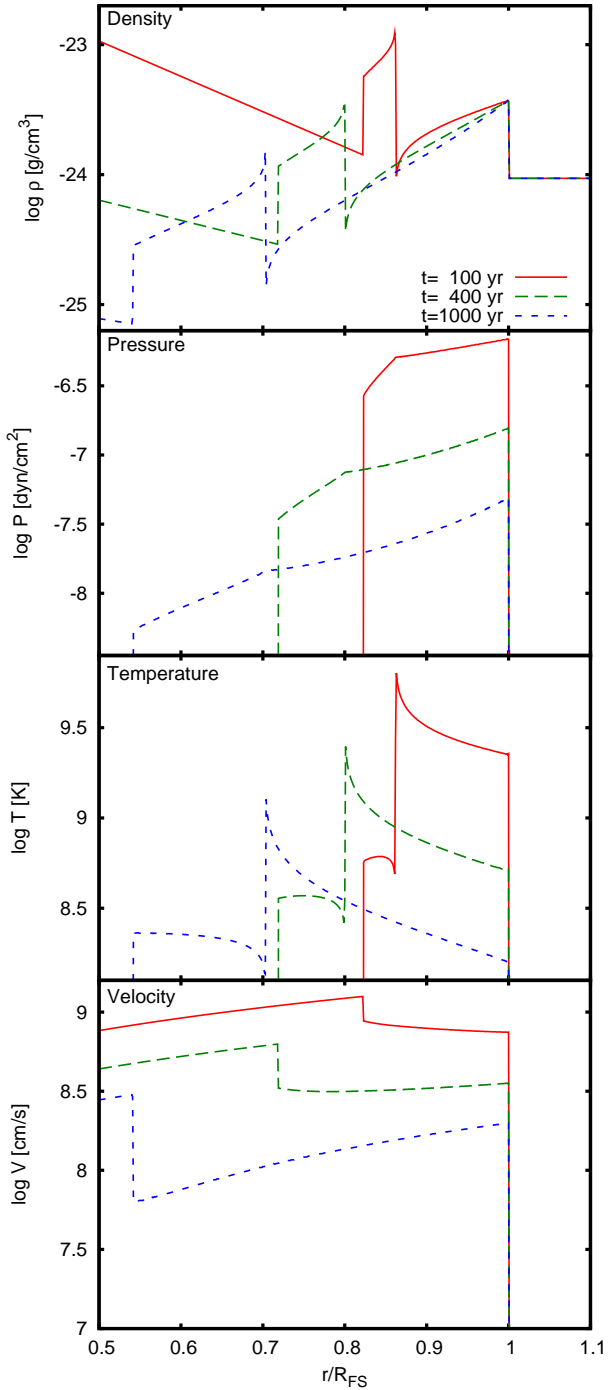}
\end{center}
\caption{Hydrodynamical parameters of the SNR at different ages. The x-axis is scaled to the radius of the forward shock.}
\label{hydropar}
\end{figure}

The initial conditions then depend on three parameters: (i) The energy
in the ejected material, which we take as the canonical energy of a SN
explosion, 10$^{51}$ erg, (ii) the mass of the ejecta, which we take to
be the Chandrashekhar mass, 1.4 M$_{\odot}$, and (iii) the density of
the surrounding medium, which we take to be 9.36 $\times 10^{-25}$ g
cm$^{-3}$, or a number density of about 0.4 for a gas with 90\% H,
10\% He, and a trace of other materials.

The supersonic expansion of the SN ejecta into the ambient medium
gives rise to a forward shock expanding into the medium, and a reverse
shock expanding back into the ejecta in a Lagrangian sense. The two
are separated by a contact discontinuity, which separates the shocked
ejecta from the shocked ambient medium. Although not captured in our
spherically symmetric simulations, in multi-dimensions the
decelerating contact discontinuity is always unstable to
Rayleigh-Taylor (R-T) instabilities, leading to R-T ``fingers'' of
shocked ejecta expanding into the shocked ambient medium.

The simulation commences with a grid of size 6 $\times 10^{-5}$ pc. By
the end of the simulation, which is carried on for about 1000 years, it
has grown to more than 10 pc. The hydrodynamic parameters at the age of
100, 400, and 1000 years are shown in Fig \ref{hydropar}. The outer shock, inner
shock and contact discontinuity are all clearly delineated as sharp
discontinuities in the density panel (top). After a few hundred years the shocked
region occupies more than 90\% of the grid. This resolution is
essential for computing the velocity differential across the shock,
required in the solution of the cosmic-ray transport equation. Note
that, while the velocity varies smoothly across the contact
discontinuity (CD), and the pressure shows a slight change in slope,
there is a dramatic change in density across the CD. The shocked
ejecta reach a maximum density on the inner side of the contact
discontinuity, while the shocked ambient medium reaches a minimum on
the outer side. The temperature therefore reaches a maximum at the CD
for the shocked ambient medium. This distinguishes it from the
structure obtained for a power-law profile expanding into the ambient
medium, as would be more appropriate for a Type II or core-collapse SN
\cite{DwaChe98}.

Finally, the numerical solution of the hydrodynamic equations is
mapped onto the spatial coordinate $x_\ast$ (cf. Eq~\ref{coord}) in
which we have written the particle transport equation
(Eq.~\ref{traneq}). The shocks, which are invariably smeared out in
the diffusive hydrodynamic calculation, are re-sharpened by
interpolation toward a point-like jump in density, flow velocity, and
pressure.  This is necessary to maintain a realistically fast
acceleration at GeV-to-TeV energies.

\section{Magnetic Field and Models Setup}
\label{modmf}
\subsection{Magnetic Field Amplification at SNR shocks}

\begin{figure}[!t]
\begin{center}
\includegraphics[width=0.49\textwidth]{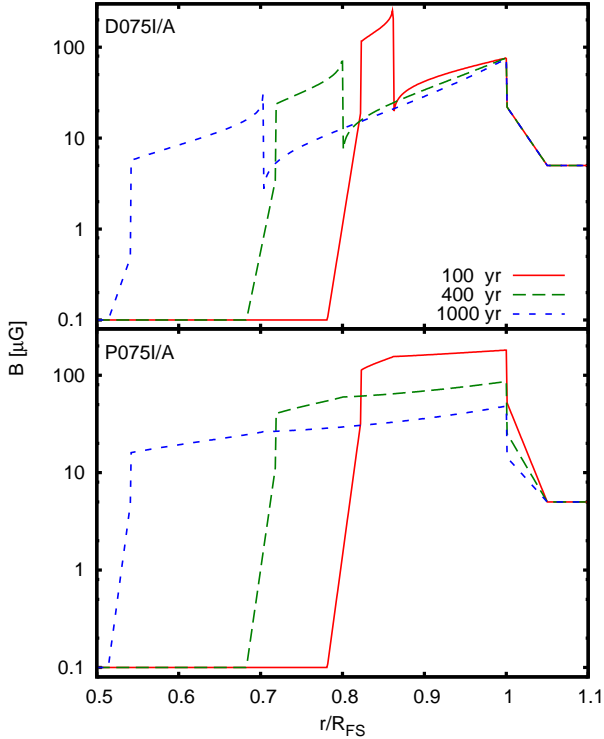}
\end{center}
\caption{Magnetic field profiles in the SNR for D and P type models at different ages.}
\label{Bevo}
\end{figure}

The strength of the magnetic field in the SNR is one of the crucial parameters for CR 
acceleration, particularly so in the shock regions.
Since we assume Bohm diffusion here, the MF determines the
mean free path of the particle before it is scattered. For higher MF
a particle will be more quickly accelerated and it can attain a higher energy before
it may escape to the far-upstream region of the forward shock. Therefore, the MF
determines the maximum energy of both protons and electrons, the latter in
addition suffering energy losses determined mainly by the same MF.

The actual mechanism of MF amplification (MFA) is a subject of ongoing
research. Despite recent progress \cite{Vlaetal06, ZirPtu08a,
    Capetal09a}, it is not yet clear what mechanism (or most probably
  a combination of mechanisms) is responsible for induction of MHD
  turbulence on scales needed for the particle acceleration (for a
  recent review see \cite{Byketal11} and references therein). Due to
  the uncertainty in the dominant mechanism of MFA, and the difficulty
  of taking into account the complexiety of the MFA process, the
  magnetic field in SNRs is often phenomenologically parametrized. If the MF is assumed to be frozen in, the
    energy density of the field may correlate with the thermal and
    relativistic particle energy densities.  Here we investigate the
  impact of widely used MF profiles on the spectra of accelerated
  particles.  We introduce a set of simple scalings which describe
the temporal and spatial evolution of the magnetic field in a SNR. The
models are listed in Table~\ref{ACmod} and are explained below in
subsection~\ref{mfpar}.

Prompted by recent high resolution X-ray observations,
  which claim to detect non-thermal emission from the reverse-shock
  region of some SNRs \cite{Gotetal01, HelVin08, Sasetal06,
    Rhoetal02}, we have chosen to consider particle acceleration at
  the reverse shock in addition to that at the forward shock.  The
  major argument against the ability of the RS to accelerate particles
  is that an exceptionally low MF exists in the ejecta, a theoretical
  argument obtained under the assumption that the magnetic flux at the
  surface of progenitor star is conserved during the expansion of the
  SNR.  However, the detection of non-thermal X-rays from the RS
  implies that there may be some processes acting against rapid
  dilution of the MF, and that a large MF at the reverse shock can
  exist. These processes may be the fluid-dynamical MHD turbulence in
  the ejecta and CD region - the initial acceleration of CRs, while
  the MF was high enough, with subsequent streaming through a confined
  ejecta region (accelerated CRs do not leave the upstream region of
  the RS as easy as that of the FS). It is also not clear whether the
  MF in the ejecta, to be scaled according to flux conservation, must
  be the field at the surface of SN progenitor.  Finally, the fact
  that shocks in SNRs are collisionless implies the existence of a
  finite MF which mediates the collisionless shock. Additional
  justification for the presence of MF at the RS can be found in
  \cite{ZirAha10}.

\begin{table}[!t]
\begin{center}
\caption{Nomenclature of MF models used in this study.}
\begin{tabular}{|l|l|l|}
\hline
\backslashbox{MF}{Profile} & Density  & Pressure \\
\hline
75$\mu$G & D075I & P075I \\
+ Alfv\'enic drift& D075A & P075A \\
\hline
300$\mu$G& D300I & P300I \\
+ Alfv\'enic drift& D300A & P300A \\
\hline
\end{tabular}
\label{ACmod}
\end{center}
\end{table}

\subsection{Parametrizations of Magnetic Field Profiles}
\label{mfpar}

The actual amplification of the magnetic field is a complicated,
  turbulent process that is not well understood. Many published
studies of particle acceleration in SNRs assume that the MF is
proportional to the density of the plasma (e.g., \cite{Ptuetal10,
  ZirAha10, Voeetal08, Beretal94}). An alternative approach is
  generally adopted in models of SNe radio emission, where the MF
energy density in the remnant is proportional to the thermal energy
density, and thereby the gas pressure \cite{1982ApJ...259..302C}
(for review see \cite{Weietal02}). Following these scalings,
we set up 8 MF models, 4 'D' models following the density profiles and
4 'P' models scaling with the square root of pressure profiles. All
our models are time dependent, but the matching at the FS is different
for the two classes of MF models.  D models assume a constant value of
the MF at the FS, $B_0 = B_{FS}$, and only the scaling inside the SNR
proportionally follows the time-dependent density distribution,
$B(r,t) = B_0 \rho(r,t)/\rho(R_{FS},t)$. In contrast, P models assume
that the MF at the FS and inside the SNR both evolve with time
according to the square root of pressure distribution, $B(r,t) =
\sqrt{C 8 \pi P(r,t)}$. Here, $\rho(r,t)$ and $P(r,t)$ are the density
and the pressure profiles of the plasma inside the SNR, and $C$ is a
proportionality constant chosen so that $B(R_{FS},t_{m}) = B_{FS}$,
where $t_{m} = 500$~years is the mid-time of simulations. We assume
two possible values of the MF at the SNR forward shock, $B_{FS} =
75$~$\mu$G and $B_{FS} = 300$~$\mu$G. These values are taken in
accordance with average limits given by the observational data on the
MF for young SNRs in X-rays \cite{Voeetal05} and gamma-rays
\cite{2011arXiv1102.3871A}. The proportionality constant, $C$, is
rather small in both cases, $C = 0.19$\% for $B_{FS} = 75$~$\mu$G and
$C = 3$\% for $B_{FS} = 300$~$\mu$G. The amplified value of
  the MF at the RS varies between $\simeq 5$ and $\simeq30$~$\mu$G
  depending on the age and model. Even if we start from a rather diluted
  MF, this requires amplification factors of $\simeq 50-300$, which are
  not unreasonable in CR-invoked or fluid-mechanical turbulent
  amplification.

The hydrodynamic profiles allow for MF scaling only inside the
shocked region. Generally one can expect that the MF amplification
processes operate throughout the entire shock precursor region
\cite{ZirPtu08a}, and thus the MF should show a smooth transition from
far-upstream values to amplified values, which are then compressed to
the given amplitudes at the shock ($B_{FS}$ in D models and
$B(R_{FS},t)$ in P models). We therefore assume that the MF falls off
to the interstellar field (5~$\mu$G) at the distance of 5\% of the FS
radius ahead of the FS, and down to a very small ejecta field
(0.01-0.1~$\mu$G) at the distance of 5\% of the RS radius toward the
interior \cite{ZirAha10}. This transition may be explained by
  an exponential decrease in the number density of high energy
  particles, which may cause MF amplification. In general, one can
  also expect that the diffusion coefficient becomes larger than Bohm
  diffusion, albeit still much lower than the Galactic one due to the enhanced flux of CRs. The transition to the Galactic diffusion
  occurs at the escape boundary, where the concentration of CRs tends to zero. Here we assume that Bohm diffusion operates up
  to escape boundary, $R_{esc}$. In our simulations, we assume that
in young SNRs the MF is fully isotropic on account of efficient
turbulent field amplification, therefore the MF compression ratio is
$\sigma_{MF} = \sqrt{(1 + 2 \sigma^2)/3} = \sqrt{11}$, where $\sigma$
is the gas compression ratio. The distributions of the MF for
different model types and different times are given at Fig~\ref{Bevo}.

\subsection{Alfv\'en Drift of Scattering Centers}

As noted earlier, we also investigate the influence on the particle
spectra of a possible drift of the scattering centers upstream of the
shocks. Resonant streaming instabilities in the precursor region of
the shock generate outward-moving Alfv\'en waves \cite{ZirPtu08a} that
may have a significant phase velocity in the presence of amplified
MF. As particle are scattered on excited MHD waves, the Alfv\'en
velocity of the waves may effectively change the compression ratio
felt by the particles, and the spectrum becomes noticeably softer
\cite{2000JPlPh..64..459L, ZirPtu08b, Capetal09a, Ptuetal10}. In cases
involving an Alfv\'enic Mach number $M_A \lesssim10$, the
test-particle slope is softer, $s < 2$, than the classical solution
\cite{KanRyu10}. It should be noted that non-resonant instabilities
often involve turbulence with very small phase velocity \cite{Bel04}
for which Alfv\'enic drift is ignorable. Which type of instability
dominates and what the turbulence properties are in the nonlinear
regime is the subject of ongoing research
\cite{2009ApJ...694..626R,2009ApJ...706...38S,2010ApJ...717.1054R}.
Here we parametrically study the impact of Alfv\'enic drift.  In half
of our models we introduce Alfv\'enic drift in the upstream regions of
both shocks assuming that the effective velocity of the scattering
centers equals the Alfv\'en velocity, $v_A = B_{FS}/\sqrt{4 \pi
  \rho_{ISM}}$, plus the upstream gas velocity. In the downstream
region, we assume that the scattering centers are isotropized and
there is no Alfv\'enic drift. We mark the models including Alfv\'enic
drift with ``A '' while the others are marked with ``I''.

\section{Results and Discussion}
\label{resad}

\begin{figure*}[!t]
\begin{center}
\includegraphics[width=0.49\textwidth]{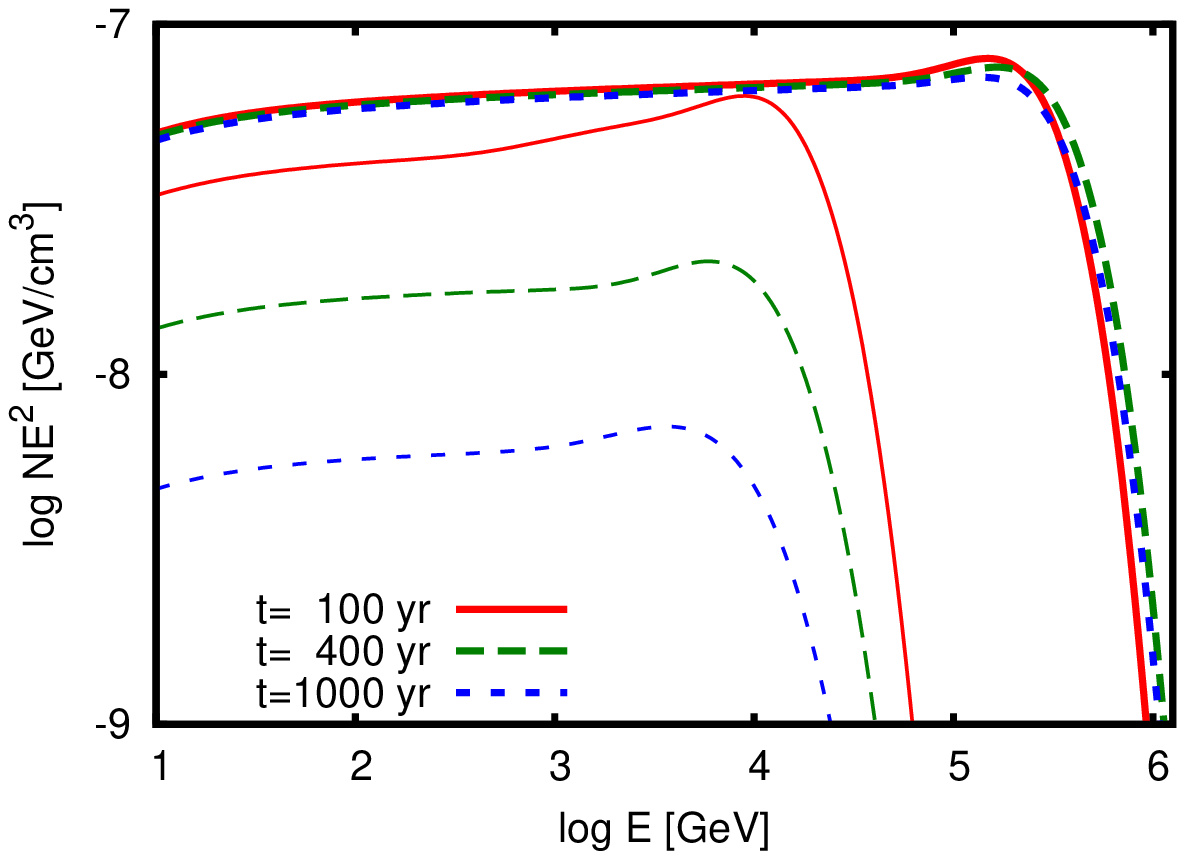}
\includegraphics[width=0.49\textwidth]{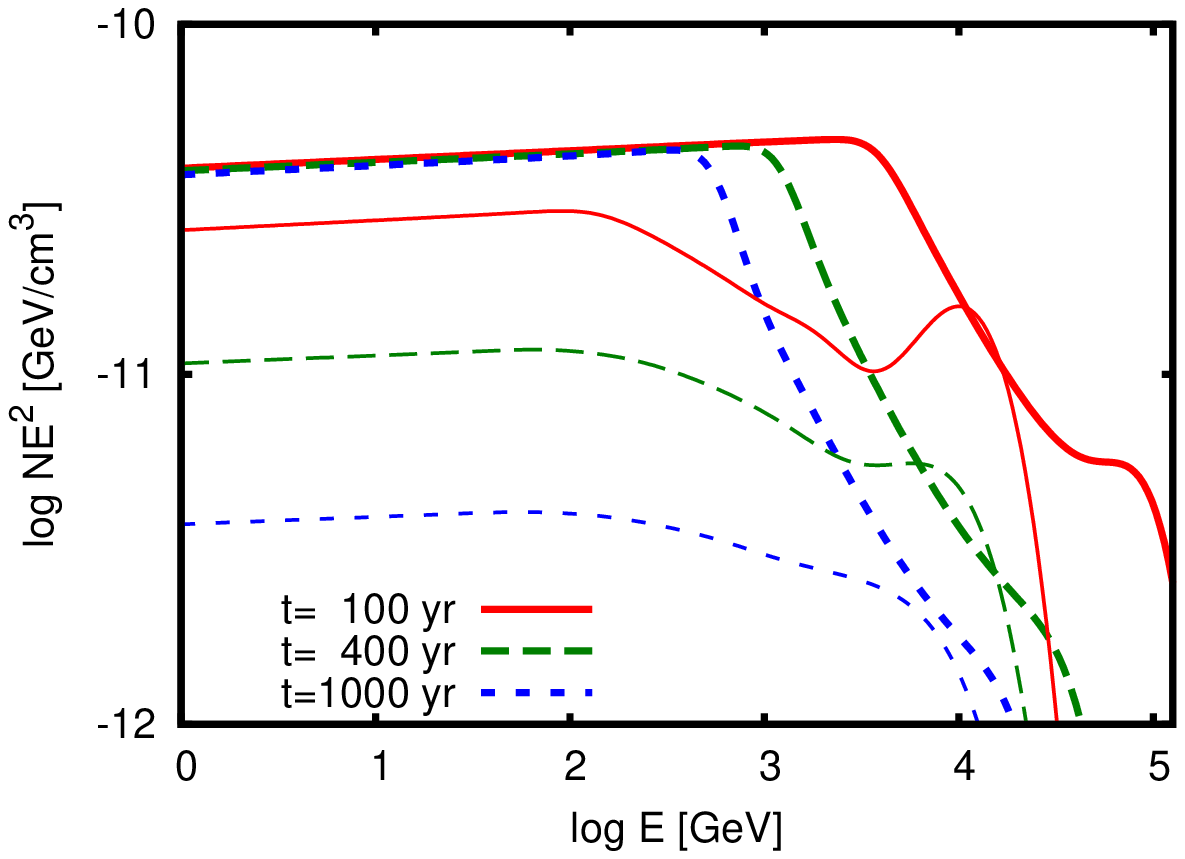}
\end{center}
\caption{The FS (thick lines) vs. the RS (thin lines) contributions to the spectra of protons (left) and electrons (right) for different SNR ages.}
\label{FSvRS}
\end{figure*}

In this section we present our main results. We discuss spectra of
particles accelerated at the reverse and the forward shocks, analyze
their properties and evolutionary differences, and explore the
intrinsic differences in particle spectra imposed by different
parametrizations of the MF and how the Alfv\'enic drift of scattering
centers affects particle spectra. Finally, we calculate the emission
from the particles in SNR and discuss the resulting observable
properties which could be compared to the experimental data.

\subsection{Particle Spectra}
\label{PS}

\subsubsection{Forward vs. Reverse Shock}
\label{FvR}

The non-thermal spectra of accelerated particles are characterized by
a power-law index and a cut-off energy. The power-law slope according
to DSA theory depends solely on the compression ratio of the plasma
flow (velocity jump across the shock). At the location of the shock,
where the plasma is compressed, particles adiabatically gain energy
and then lose it everywhere in the expanding flow. In the evolving
system, the average compression ratio should be considered. Our
hydrodynamical simulations were performed with sufficiently high
resolution that the deviations of the shock compression ratio do not
affect the spectral slope through the entire evolution. The
compression ratios in our simulation are $\sigma_{FS} = 4.01\pm0.09$
and $\sigma_{RS} = 3.97\pm0.11$, for the FS and the RS respectively.

In the models without Alfv\'enic drift, the particles accelerated at
the FS obey the classical simple power-law distributions with index
$s\simeq 2$ up to the energy, at which the escape of the particles
from the system (for protons) or the energy losses (for electrons)
inhibit further acceleration. As predicted by DSA \cite{Dru83}, for
protons $E_{max,p} \sim V_{sh}^2 B_0 t$. Therefore, the increase of
$E_{max,p}$ due to system expansion (higher energies are needed to
leave the system) is eventually compensated by the decrease of
$V_{sh}$, so $E_{max,p}$ decreases after reaching a maximum. This is
illustrated by the time evolution of FS proton spectra in
Fig.~\ref{FSvRS} (left), which shows the D300I model, in which $B_0$ is
kept constant, thus isolating $E_{max,p}\propto V_{sh}^2 t$. For
electrons, $E_{max,e}$ evolves in the same manner up to the time when
the energy gain per cycle no longer exceeds the energy loss per cycle
due to synchrotron radiation, i.e, $t_{acc} = t_{rad}$. A second,
typically lower, characteristic energy, $E_{rad}$, is approximately
set by equality of the radiative-loss timescale with the age of the
remnant. Above $E_{rad}$, the electron spectra assume a quasi
steady-state, changing slowly on account of the evolution of the
gas-flow profiles. Besides, above $E_{rad}$ the electron spectra
steepen by one power approximately (Fig.~\ref{FSvRS}, right).

The ability of the RS to accelerate particles depends on the MF
configuration at the location of the RS. With the MF profiles assumed
here and discussed above, it is observed that the RS is
capable of accelerating particles to sufficiently high energies and
intensities to contribute to the total volume-integrated particle
spectra. These results are at least consistent with many of the
  observations noted above. In order to obtain the volume-integrated
spectra, $N(E)$, one performs an integral of the spatially
differential particle distribution, $N(x,E)$, over the entire volume
of the simulation grid up to $R_{esc} = 2 R_{FS}$:
\begin{equation}
N(E) = 4 \pi R_{FS}^3 \int_0^2 N(x,E) x^2 dx
\end{equation}
where $x=r/R_{FS}$. In all plots featuring particle spectra we show the
quantity $N = N(E)/4 \pi R_{FS}^3$. In the absence of Alfv\'enic
drift, the RS spectra show slopes similar to those at the FS, with the
exception of some hardening at higher energies.  The maximum energies
for protons are lower than at the FS on account of a lower MF
strength, which allows particles to escape the acceleration region at
lower energies. In contrast, the maximum energies for electrons are
nearly as high as at the FS mainly because a weaker MF also reduces
the rate of radiative losses of electrons. The intensity of the
volume-integrated RS spectra falls with time relative to the FS
spectra as the number of injected particles becomes smaller. Also the
ratio of the RS surface to FS surface decreases with time, which means
fewer particles that could be potentially injected cross the RS.

\begin{figure*}[!t]
\begin{center}
\includegraphics[width=0.98\textwidth]{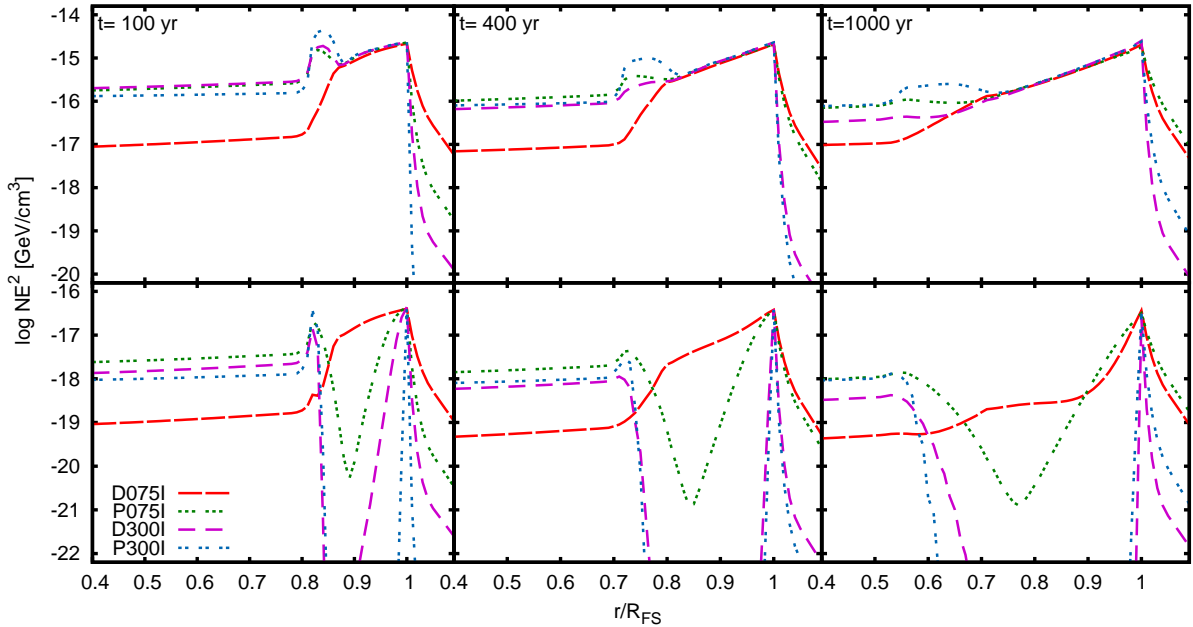}
\end{center}
\caption{The time evolution of the spatial distribution of protons
  (top row) and electrons (bottom row) at energy of 20~TeV.}
\label{mfi}
\end{figure*}

The stronger spectral hardening and more significant
bump at high energies make the RS volume-integrated spectra different
from those of the FS. The spectral hardening is most visible at early
evolutionary epochs and gradually disappears with time for both
shocks, but at a somewhat faster rate for the FS. The bump
in the forward and the reverse shock spectra is located close
to $E_{max}$, which is beyond $E_{rad}$ for electrons. The presence of
these spectral features and their evolution can be explained by the
geometry of our spherically-symmetric system.

In order to get volume-integrated spectra we integrate over
  the entire simulation domain, so both upstream and downstream
  regions contribute to the volume-integrated spectra.  The
volume-integrated upstream spectra usually have hard slopes, $s \simeq
1$, and the intensity of the upstream spectra around $E_{max}$ is
comparable with that of the downstream spectra. Thus, in spectra
integrated over both upstream and downstream regions one can see the
contribution from the hard upstream spectra in high-energies close to
$E_{max}$, whereas the contribution of low-energy particles is
negligible. Besides, the $E_{max}$-region of the upstream spectra
exhibits a bump which appears on account of the exponential decrease
of the MF in the upstream direction. As a result, the diffusion time
of particles located close to the shock is longer than that of
particles further away from the shock. Therefore, the recently
accelerated particles around $E_{max}$ are accumulated near the shock
in the upstream region. The spectral hardening and the bump is
strongly visible in the RS spectra (Fig.~\ref{FSvRS}, right)
because the volume of the upstream regions of both shocks are
  significantly different in the spherically-symmetric geometry
  assumed here, contrary to the case of plane-parallel shocks. This
  also explains the different time evolution of the spectral features. The
  upstream region of the FS expands significantly faster than the
  upstream region of the RS, thus the CR number density in the upstream
  region of the FS decreases faster and contributes less to the total
  volume-integrated spectra. We note the presence of a peculiar wiggle
  in the electron spectra. The MF in the upstream region is lower
than downstream, resulting in weaker synchrotron losses for the
electrons and therefore $E_{rad} \simeq E_{max,e}$. Consequently, the
upstream electron spectra have a significantly higher intensity near
$E_{max,e}$ than the downstream electron spectra. Therefore, in the
total volume-integrated spectra the bump stands out beyond $E_{rad}$
in the $E_{max,e}$ region. For protons, $E_{max,p}$ in the upstream
and downstream spectra nearly coincide, and a small bump appears near
$E_{max,p}$ in the total volume-integrated proton
spectra. Again, due to the faster decrease in the CR number density
  ahead of the FS with time, the feature is more prominent in the RS
  spectra.

\subsubsection{Influence of Magnetic Field Profiles and Strengths}
\label{MFI}

Most important for DSA is the configuration, strength, and evolution
of the MF at the location of the shocks and their precursor
regions. Since here we adopted similar profiles of MF in the precursor
regions of the shocks, only a different time evolution of the MF
strength at the location of the shocks may change the shape of the
volume-integrated spectra. It was noted, however, that the differences
between models in the spectra of particles created by the RS and the
FS are marginal, in fact only $E_{rad}$ and $E_{max}$ truly depend on
the MF strengths. While for protons a higher MF would mean a higher
$E_{max,p}$, for electrons this would mean a lower $E_{rad}$ and
$E_{max,e}$ due to radiative losses.

The MF profiles in the SNR interior affect the radial distribution of
CR particles and their subsequent emission, especially so for
leptons. In Fig.\ref{mfi} we plot the evolution of the radial
distribution of protons and electrons at the energy of 20~TeV, which
could be a characteristic energy of the particles producing
high-energy emission, discussed in more detail in
subsection~\ref{BRP}. We use models without Alfv\'enic drift as it
does not change the distribution significantly - it would only
slightly reduce the intensity at the FS. One can clearly see that
electrons suffer significant losses in the shocked region where the MF
is high. The number density of protons falls nearly exponentially from
the FS towards the CD. There is a noticeable peak in the CR
distributions near the RS. Whereas electrons peak exactly at the RS,
protons are accumulated between the RS and the CD. At 100 years, the
CR number density close to the RS is even higher than at the FS. At
400 years the peak is still clearly visible, however already at 1000
years only a minor bump is observed. In the ejecta region, where the
MF is low, both particle species show nearly uniform
distributions. Although the D075I model follows the trends described
above, it is a clear outlier. The cut-off energy of particles
accelerated by the RS in this model is around 4~TeV, so one finds at
20~TeV almost no particles from the RS. This explains the absence of a
peak near the RS and the low CR number density in the interior of the
SNR. Besides, the electrons have suffered little energy losses in the
initial phases of SNR evolution ($t < t_{rad}$) compared with other
models (remember that P075I model have initially a high MF).

\begin{figure}[!t]
\begin{center}
\includegraphics[width=0.49\textwidth]{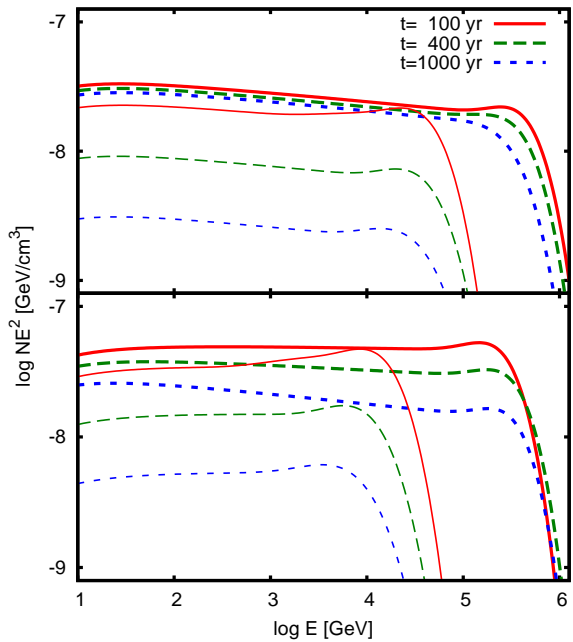}
\end{center}
\caption{The spectra of the FS (thick lines) and the RS (thin lines)
  for different SNR ages in P300A (top) and D300A (bottom) models.}
\label{alf}
\end{figure}

\subsubsection{Impact of Alfv\'enic Drift}
\label{ALF}

We have explored the influence of Alfv\'enic drift on the particle
spectra of the SNR. Interesting is that it affects our models in a
different way during the SNR evolution. We observe that the FS and the
RS spectra in P models are noticeably softer during all evolutionary
times. This is because in P models the Alfv\'enic Mach number, $M_A$,
is almost constant for both shocks during the evolution. The MF
strength, $B$, evolves as the square root of the pressure, $\sqrt{P}$,
and hence roughly scales with the shock velocity. Therefore, spectra
show nearly the same softening at all times as shown at the top of
Fig.~\ref{alf}. Contrary to P models, in D models the effect of
Alfv\'enic drift is different for the FS and the RS. At the FS $B$ is
kept constant during the evolution, and $M_A$ is very large at early
times because the Alfv\'en velocity, $v_A$, is small compared to the
shock speed. At later times $v_A$ remains constant, but the shock
speed drops, and so $M_A$ becomes small. This implies that the
softening of the FS spectra in D models was small at early times and
increased during the evolution of the SNR. In contrast to the FS, $B$
is evolving at the RS as density, $\rho$. The initially large ejecta
density caused a high Alfv\'en velocity ($v_A \sim \sqrt{\rho}$ in D
models) and rather small $M_A$, though larger than at the FS, which
caused some softening of the spectra. Over time $\rho$ dropped
significantly, and therefore $M_A$ became large. Therefore, the
softening of the RS spectra in D models is stronger at earlier times
and almost not noticeable at later times (Fig.~\ref{alf},
bottom). Obviously, all these effects and characteristics are stronger
in models with high MF strength.

\subsubsection{Total spectra from both shocks}
\label{SUM}

\begin{figure*}[!t]
\begin{center}
\includegraphics[width=0.98\textwidth]{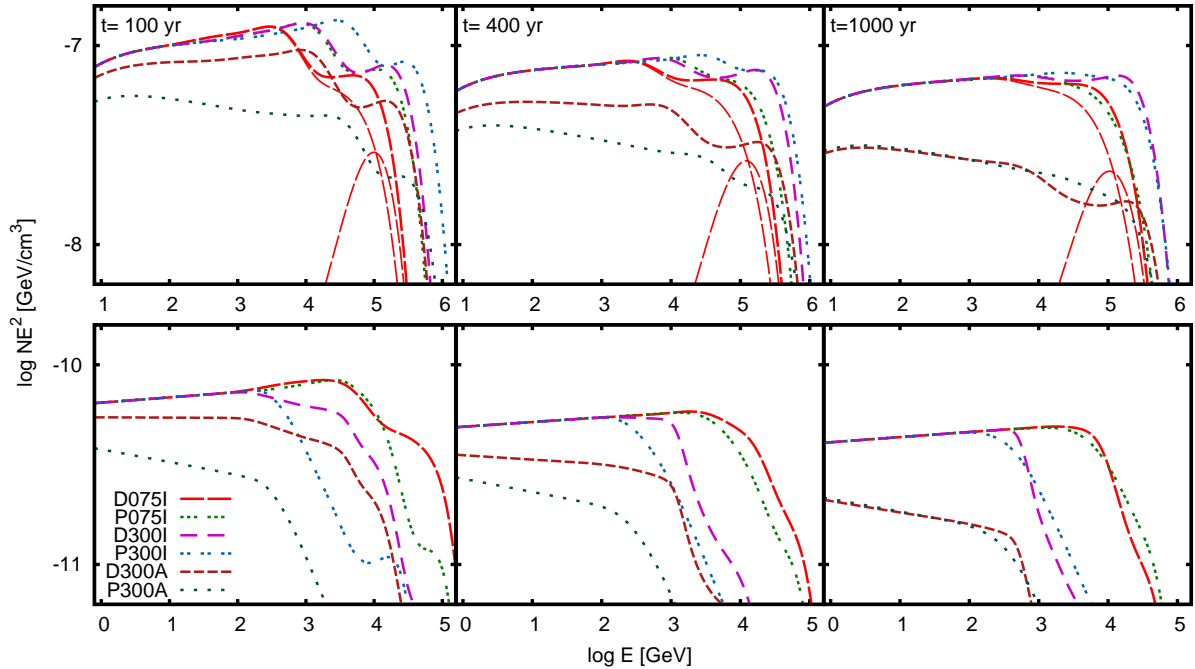}
\end{center}
\caption{The time evolution of volume-integrated proton (top row) and
  electron (bottom row) spectra for different models and SNR ages. Thin lines for the D075I model in the top row display the downstream and upstream components of the proton spectra.}
\label{sum}
\end{figure*}

The shape of the total volume-integrated spectra depends on the SNR
age on account of the changing importance of the contributions from
the forward and reverse shocks. For very young Type Ia SNRs, and
  keeping in mind the MF profiles assumed herein, a significant
contribution to the total spectrum arises from the RS. However,
already after a few hundred years the bulk of particles is provided by
the FS. Nevertheless, particles accelerated at the RS continue to
  create peculiar spectral features up to late times of the SNR free
expansion phase (see Fig.~\ref{sum}). Added to this is the different
impact of Alfv\'enic drift on the spectra produced with different MF
models and the time evolution of this impact, especially in D
models. The total spectra are very different from a single-shock
test-particle solutions as seen at Fig.~\ref{sum}. We plotted 6 out of
8 models because D075A and P075A models show similar shapes as D075I
and P075I models, apart from being slighter softer.

At 100 years the contribution of the RS to the total spectra
  is significant in all models. With age this contribution
decreases. After 400 and 1000 years, the proton spectra in P models
look like broken power laws or vary gradually, much smoother
  than exponentially cut-off. Since the RS proton spectra in D models
  have a somewhat smaller cut-off energy and are harder than the FS
  spectra, if Alfv\'enic drift is assumed, the RS maintains a
  significant contribution to the total spectrum until late
  epochs. Because of it and because of the bump contributed by the
  upstream region of the FS, the proton spectra of D models show some
  small-scale concavity at high energy through all times, however at
  later times it is shallower and broader (bumps in both RS and FS
  spectra become less significant).

The total electron spectra (Fig.~\ref{sum}, bottom row) also have
interesting features. On account of energy losses, the contribution of
the RS in the cut-off region is stronger. This creates either
small-scale concavity or a broken power-low in the loss-affected
region (between $E_{rad}$ and $E_{max,e}$). The stronger the
contribution from the RS is, the more pronounced a concavity is
observed. When the contribution of the RS is mild, then one sees a
broken power-law. In some models, (i.e., D300I and D300A), the shallow
concavity in the loss-affected region is visible throughout all late
times. It is striking that the electron spectra in single-shock NDSA
simulations \cite{Kan11} have similar concavity in the loss-affected
spectral band. Additionally, one can note that near $E_{rad}$ the
spectra of P models are smoother than the spectra of D models.

We must make an important remark concerning particle
  spectra. The plotted spectra are integrated over the entire
  simulation domain. However, one finds (\ref{RSP}) that the main
  contribution to the emission is done by the particles confined
  inside the remnant. These particle spectra are typically softer at
  high energies because they do not include the hard upstream
  particles. For illustration see the thin lines of D075I models in
  Fig.~\ref{sum}, which show separately SNR-confined (downstream)
  particles and peaked at high energies upstream particles. Only
  inverse Compton emission because of low MF in the upstream region
  can benefit from the upstream particle population.

\subsection{Particle Radiation}
\label{PR}

\begin{figure*}[!t]
\begin{center}
\includegraphics[width=0.98\textwidth]{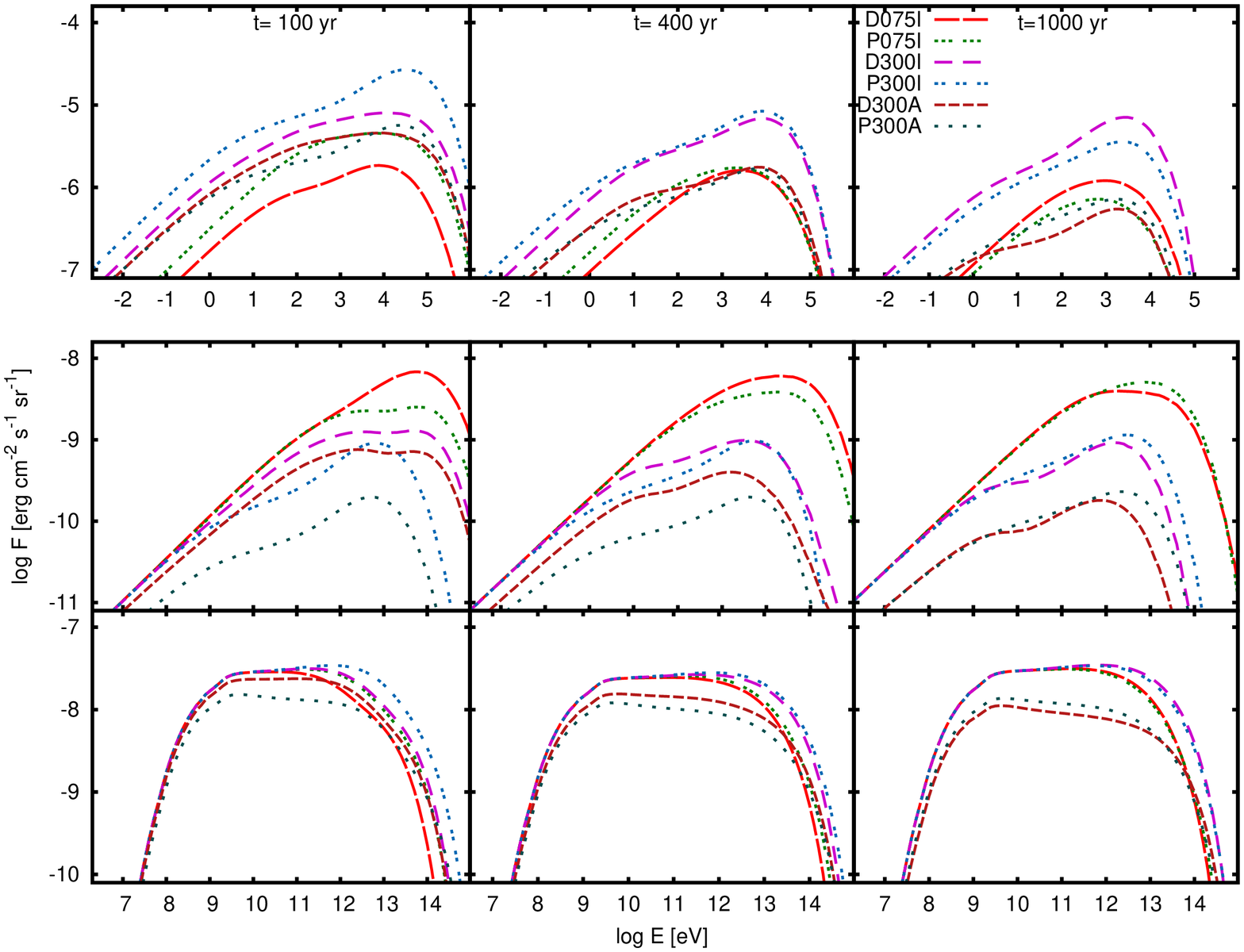}
\end{center}
\caption{The time evolution of volume-integrated emission from the SNR
  due to synchrotron (top row), inverse Compton (middle row) and
  pion-decay (bottom row) for different models and SNR ages.}
\label{rad}
\end{figure*}

We consider three basic non-thermal emission processes, synchrotron
and inverse Compton (IC) emission of electrons, and neutral pion
decays originating in collisions of CR protons with protons at
rest. The synchrotron emission is calculated according to
\cite{Stuetal97}. We consider only the CMB photon field for the IC
scattering. Therefore, we may use a simple rescaling of the
synchrotron emission to IC scattering \cite{Poh96, Aha04}. We
calculate the gamma-ray emission produced by pion decays according to
the procedure described in \cite{Huaetal07}. Thermal emission is
not considered. However, we made estimates that even assuming instant equilibration the maximum of the thermal emission does not
exceed 10\% of the synchrotron emission at the respective energy for
all considered ages. In particular, at 3~keV the thermal component may
comprise $\simeq 3$\% of the X-ray emission of the models considered
here. The reader should be aware though that we used the same
injection efficiency for electrons and protons which results in high
amplitudes of leptonic emission. If the electron injection
efficiency is lowered significantly, below 3~keV the
brightness profiles may be affected.

\subsubsection{Radiation Spectra}
\label{RSP}

The radiation spectra show diversity during SNR evolution (see
Fig.~\ref{rad}). In general, the radiation spectra are much smoother
than the parent particle distributions. The leptonic spectra show
broad features, which are the result of energy losses experienced by
electrons during their acceleration history. A noticeable feature of
the leptonic spectra is concavity, which appears in different models
at different times. At 100 years it is strongly visible in P300I and
P300A models, but in D075I and D075A it is nearly invisible. Later, at
400 years, it appears in D300I and D300A models while it disappears in
D075I and D075A, and it starts to vanish in P300I and P300A models. At
1000 years it is visible only in D300I and D300A models. The concavity
reflects the change in electron slope on account of the contribution
of the RS between $E_{rad}$ and $E_{max,e}$. It interesting to see
that the volume-integrated synchrotron and IC spectra have different
shapes. While the synchrotron emission comes mainly from the shocked
region, a significant fraction of IC emission comes from the ejecta
and upstream region with low MF, where the electron spectrum
is different. The pion-decay spectra are much smoother than their
leptonic counterparts and can be characterized by a power-law with a
very gradual and broad cut off at relatively low energy between a few
100 GeV to a few TeV. This makes the spectra rather soft in the energy
band observed with modern Cherenkov telescopes. The pion-decay
  emission comes mainly from inside the SNR, with a small amount of upstream
  contribution.

\subsubsection{Brightness Profiles}
\label{BRP}

\begin{figure*}[!t]
\begin{center}
\includegraphics[width=0.98\textwidth]{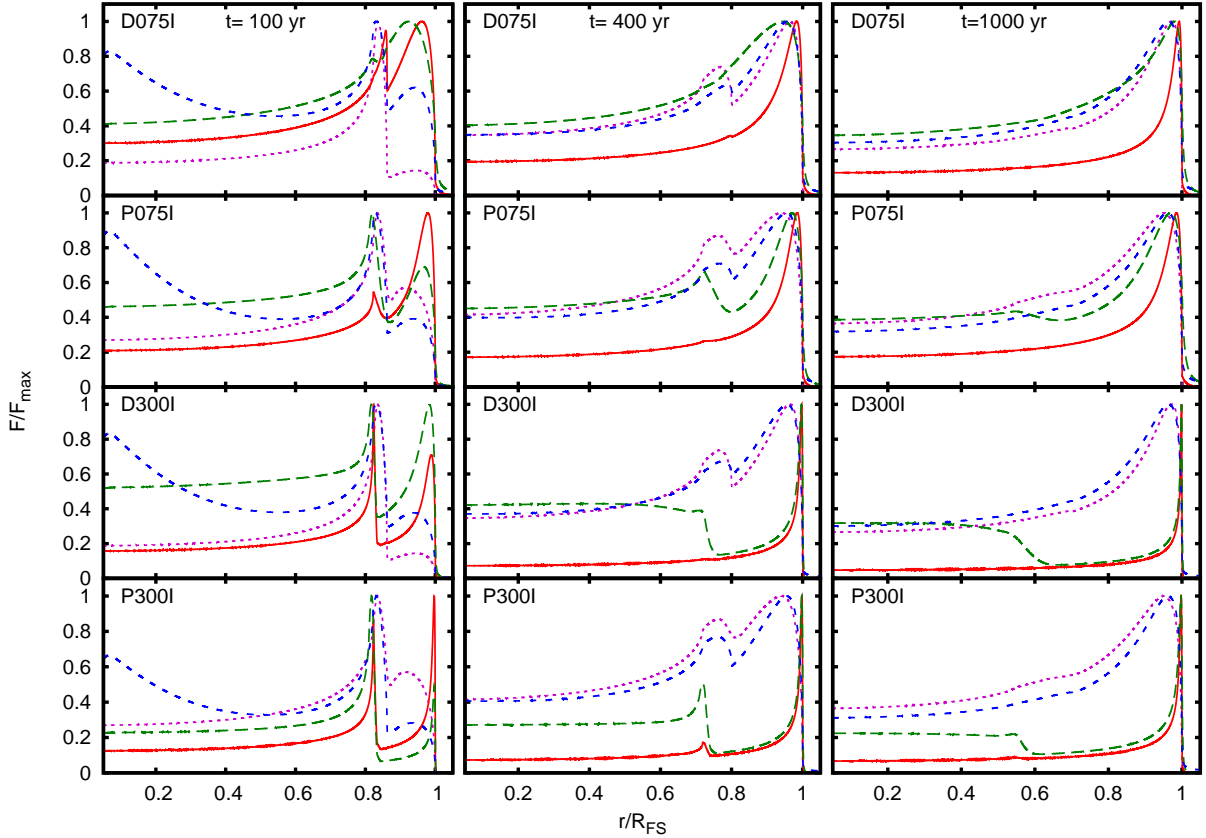}
\end{center}
\caption{The radial intensity profiles of the SNR in different models
  at different wavelength and ages. Radio @1.4~GHz (dotted-pink),
  X-rays @3~keV (solid red), IC @1~TeV (long-dashed green), pion-decay
  @1~TeV (dashed blue).}
\label{pro}
\end{figure*}

To permit morphological studies of SNR at different ages, we computed
radial intensity profiles at characteristic wavebands for current
instrumentation (radio @1.4~GHz, X-rays @3~keV, gamma-rays @1~TeV). In
Fig.~\ref{pro} we separately plot the emission of leptonic and
hadronic origin. The profiles are normalized to the emission maximum
along the SNR radius. The absolute intensity may be obtained from the
spectra given at Fig.~\ref{rad}. We plot here ``I'' models only. Their
``A'' counterparts show similar profiles, but the intensity of the FS
is a bit lower, making the profiles look somewhat flatter.

It is astonishing to see how the brightness profiles undergo severe
changes with time. At 100 years both shocks are bright and visible in
all bands except for the FS in radio waves, which are emitted by
low-energy electrons. The contribution of the RS to the total emission
is high at the initial stages. The low-energy electrons cannot
propagate far from their point of origin, and therefore the relatively
high injection rate at the RS in the early phase renders it much
brighter than the FS in radio band. Pion-decay gamma rays are
copiously produced in the ejecta region on account of the high target
density there. At 400 years the contribution of the RS region becomes
less prominent, but is still visible. Only in synchrotron X-rays (and
IC in D075I/A models) the RS becomes insignificant. The 3~keV
synchrotron X-rays are created by electrons of rather high energy
($\sim 13.5$~TeV for $B=300$~$\mu$G and $\sim 27$~TeV for
$B=75$~$\mu$G). At this age the RS is no longer able to boost
electrons up to these energies. The electrons accelerated at the FS
suffer from losses on their way to the SNR interior, and therefore the
synchrotron emission becomes noticeably filamentary, especially in
high-MF models. A characteristic feature develops in IC emission of
the models with high MF. A plateau of high intensity appears in the
ejecta region and a low intensity between the two shocks. In the
shocked region, high-energy electrons are few on account of severe
energy losses in the high MF, so the IC emission is suppressed. At the
same time, the electrons that propagated into the ejecta region, where
the MF is low, accumulate there and account for bright IC
emission. This trend continues with age, and after 1000 years we
observe a bright plateau of IC emission from the SNR center, while the
FS appears very filamentary. There is almost no trace of emission from
the RS after 1000 years in the other wavebands. X-ray synchrotron
emission is seen as a thin filament near the FS, especially in the
models with high MF.

\section{Conclusions}
\label{concl}

We have studied particle acceleration by both forward and reverse
shocks in young type-Ia Supernova Remnants. Rather than starting
  with Sedov models, as is most commonly done, we used gas-flow
  profiles in the ejecta-dominated phase, derived from 1-D
hydrodynamical simulations, to solve the cosmic-ray transport equation
in a test-particle approach. We analyzed how different
phenomenological parametrizations for the magnetic-field
distribution and strength affect the maximum energy and spatial
distributions of accelerated particles. Additionally, we have explored the
influence of possible Alfv\'enic drift of scattering centers on the
total particle spectra in the upstream region of both SNR shocks
and find our results for the forward shock in agreement with
  previous findings \cite{ZirPtu08b, Capetal09a, KanRyu10}. We
investigated the properties of the resulting non-thermal radiation,
i.e., its spectra and the spatial distribution of the intensity at
characteristic wavelengths.

We have demonstrated that it is important to account for the
contribution of the RS to cosmic-ray particle population in the
initial 400 - 600 years of SNR evolution, given the widely
  used parametrizations of the MF that are assumed here. At that time
the RS is able to accelerate particles up to very high energies, and
the number of accelerated particles is comparable to that at the
FS. This is well visible in volume-integrated particle spectra, the
distribution of high-energy particles in the SNR, as well as in the
morphology of particle radiation. The significance of the RS
contribution falls with time. Although the RS contribution is still
visible in particle spectra up to 1000 years or so, it is barely
noticeable in the emission spectra and morphology, which is in
  agreement with observations.

We found that the total volume-integrated particle spectra of the SNR
can be very different from single-shock test-particle solutions,
especially for very young SNRs. The choice of MF profiles affected the
relative contribution to the total spectrum of the RS and the FS, and
consequently the emission spectra in the high-energy region
varied. Additionally, Alfv\'enic drift of scattering centers in the
shock precursors gives rise to additional spectral feature, whose
appearance and time evolution depends on the choice of MF
profile. Generally, P models are more sensitive to Alfv\'enic drift
than D models. Obviously, the models with high MF are affected
stronger than those with low MF, for which the effect is subtle.

The particle distributions computed here show a variety of spectral
shapes. Interesting is that some of the spectra (both electron and
protons) exhibit features which may be considered similiar to
  ones shown using NDSA, such as spectral concavity and high-energy
bumps. However, concavity in our spectra arises in a much smaller
  high-energy band compared to the broadband concavity of NDSA. Besides,
  it is not reflected in pion-decay spectra as these are produced
  predominantly by confined particles in the SNR. Additionally, our
test-particle spectra are softer at high energies than those of NDSA,
even if both are modified by Alfv\'enic drift, which may provide
better agreement with gamma-ray spectra observed from SNRs.



%% References

\newpage
\newpage
\bibliographystyle{elsarticle-num}
\bibliography{snrcr}






%\end{thebibliography}

\end{document}